\documentclass[a4paper, oneside, twocolumn, notitlepage, 10pt]{extarticle_ecoc}
\usepackage{ecoc}

\usepackage{pgfplots}
\pgfplotsset{compat=1.18}
\usepgfplotslibrary{groupplots}
\definecolor{uclpurple}{RGB}{154, 59, 255}
\definecolor{uclblue}{RGB}{36, 136, 242}
\definecolor{uclgreen}{RGB}{47, 148, 26}
\definecolor{uclorange}{RGB}{255, 118, 38}
\definecolor{uclred}{RGB}{229, 28, 0}

\definecolor{uclwhite}{RGB}{250, 250, 250}
\definecolor{ucldpurple}{RGB}{54, 26, 84}    \definecolor{uclbpurple}{RGB}{153, 59, 255}  \definecolor{uclmpurple}{RGB}{186, 130, 255} \definecolor{ucllpurple}{RGB}{221, 189, 255} \definecolor{uclppurple}{RGB}{238, 222, 255} \definecolor{uclhblue}{RGB}{48, 214, 255}    

\definecolor{uclg0}{RGB}{54,26,84}    \definecolor{uclg1}{RGB}{153,59,255}  \definecolor{uclg2}{RGB}{0,46,166}    \definecolor{uclg3}{RGB}{120,28,28}   \definecolor{uclg4}{RGB}{158,26,84}   \definecolor{uclg5}{RGB}{0,94,92}     \definecolor{uclg6}{RGB}{84,135,255}  \definecolor{uclg7}{RGB}{245,99,0}    \definecolor{uclg8}{RGB}{237,54,125}  \definecolor{uclg9}{RGB}{0,158,156}   

\pgfplotscreateplotcyclelist{uclcycle_bright}{
    {uclg1},
    {uclg6},
    {uclg7},
    {uclg8},
    {uclg9}
} 
\begin{document}
\selectlanguage{english}

\title{One Terahertz Full-Field Digital Back-Propagation over 3000 km}

\author{
    Eric~Sillekens\textsuperscript{(1,\textdagger)},
    Ruben~S.\,Luis\textsuperscript{(2)},
    Giammarco~Di~Sciullo\textsuperscript{(3)},
    Robert~Emmerich\textsuperscript{(4)},
    Carlo~Centofanti\textsuperscript{(3)},\\
    Daniele~Orsuti\textsuperscript{(5)},
    Robson~A.\,Colares\textsuperscript{(6)},
    Mindaugas~Jarmolovi\v{c}ius\textsuperscript{(1)},
    Ronit~Sohanpal\textsuperscript{(1)},\\
    Darli~A.\,A.\,Melo\textsuperscript{(6)},
    Luca~Palmieri\textsuperscript{(5)},
    Colja~Schubert\textsuperscript{(4)},
    Ronald~Freund\textsuperscript{(4)}, \\    
    Cristian~Antonelli\textsuperscript{(3)},
    Robert~I.\,Killey\textsuperscript{(1)},
    Polina~Bayvel\textsuperscript{(1)}, and
    Hideaki~Furukawa\textsuperscript{(2)}
}

\maketitle

\begin{strip}
    \begin{author_descr}

        \textsuperscript{(1)} Optical Networks Group, UCL (University College London), Gower Street, London, WC1E 6BT, UK

        \textsuperscript{(2)} NICT, Nukui Kitamachi 4-2-1, Koganei, 184-8795 Tokyo, Japan

        \textsuperscript{(3)} Università degli Studi dell'Aquila, 67100, L’Aquila, Italy

        \textsuperscript{(4)} Fraunhofer Heinrich-Hertz-Institut, Einsteinufer 37, 10587 Berlin, Germany
        
        \textsuperscript{(5)} Università degli Studi di Padova, Via G. Gradenigo 6/B, 35131, Padova, Italy

        \textsuperscript{(6)} Universidade Estadual de Campinas (Unicamp), Campinas, SP, 13083-859, Brazil

        \textsuperscript{(\textdagger)}\textcolor{blue}{\uline{e.sillekens@ucl.ac.uk}}

    \end{author_descr}
\end{strip}

\renewcommand\footnotemark{}
\renewcommand\footnoterule{}

\begin{strip}
    \begin{ecoc_abstract}
We implement full-field digital back-propagation with a 1-THz receiver using 20 synchronous frequency-adjacent coherent receivers with digital stitching and a frequency-comb local oscillator. Relative to electronic dispersion compensation, per-channel DBP and full-field DBP achieve throughput gains of 2.2\% and 5.4\%, respectively. ©2026 The Author(s)
    \end{ecoc_abstract}
\end{strip}

\section{Introduction}

\begin{figure*}[b!]
    \centering
    \includegraphics[width=\linewidth]{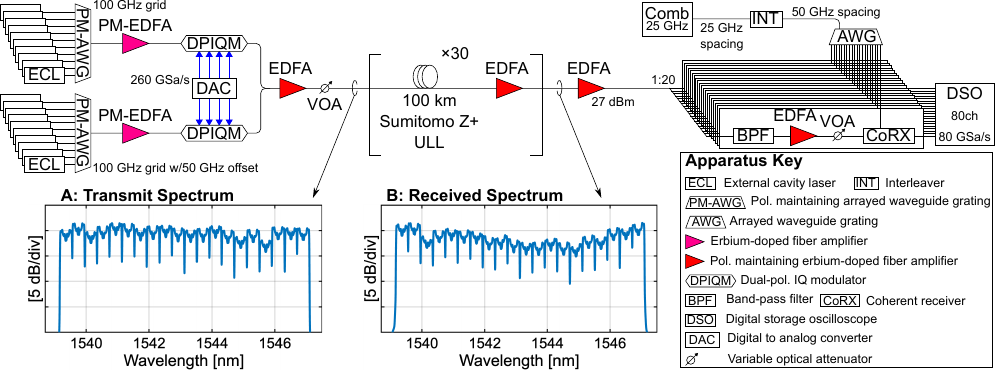}
    \caption{Experimental setup for the 1-THz full-field DBP demonstration. Two DPIQMs are used to modulate odd and even channels, which are then combined and launched into the straight-line link. At the receiver, the signal is split into 20 branches, each of which is filtered and fed into a coherent receiver, with filtered lines from a comb used as local oscillators. The transmitted and received spectra are shown in insets (A) and (B), respectively.}
    \label{fig:setup}
\end{figure*}

Full-field digital back-propagation (DBP) is the process by which all interfering signals in a transmission link are jointly detected and incorporated into a wideband DBP computation \cite{ip2010Nonlinear}. This allows the compensation of inter-channel fibre nonlinearities, such as cross-phase modulation and four-wave mixing, leading to higher nonlinearity-compensation gains \cite{dar2016Limits, goroshko2015Fundamental}. In order to capture a wideband spectrum, it may be necessary to frequency-stitch signals detected by multiple synchronous coherent receivers, as shown in \cite{fontaine2012228GHz, shi2017246a, deakin2024bandwidth, drayss2023Nonsliced,drayss2025Optical,civelli2025Twist}. It may also be necessary to use frequency comb as local oscillator, in order to achieve a consistent frequency spacing between receivers or transmitters \cite{sohanpal2023Impact}. At sufficiently wide bandwidths, third-order dispersion ($\beta_3$) must also be included in the propagation model, since the dispersion slope across 1~THz amounts to the equivalent of more than 100~km of dispersion after 3000~km of fibre \cite{ali2018Impact}. Nevertheless, the gain of full-field DBP may be fundamentally limited by phenomena such as noise \cite{galdino2017limits} or polarisation-mode dispersion \cite{czegledi2017PMD}. As such, further research is required to evaluate the potential of this approach to support long-distance transmission.

In this work, we demonstrate full-field 1-THz DBP on a 20$\times$47.5~GBd, 50-GHz-spaced, PM-64QAM WDM signal after transmission through a 3000-km straight-line pure-silica-core G.654 fibre link. The full-field receiver is implemented using 20 synchronous coherent receivers, each with 36~GHz electrical bandwidth and 11~GHz overlap for coherent stitching, and 50-GHz-spaced coherent laser lines from an optical frequency comb as local oscillators. Full-field DBP achieved a 393-Gbps (5.4\%) gain over EDC, more than twice the 160-Gbps (2.2\%) gain achieved by per-channel DBP.

\section{Experimental Transmission Setup}

A diagram of the experimental setup is shown in Figure~\ref{fig:setup}. At the transmitter, 20 free-running external-cavity lasers (ECLs) with linewidths below 100~kHz were separated into odd and even sub-bands and combined using two 100-GHz arrayed-waveguide gratings (AWGs) with a relative 50-GHz offset. Each group of sub-bands was modulated independently using a dual-polarisation in-phase/quadrature modulator (DPIQM), after which the two groups were recombined with a 3-dB coupler and amplified by an erbium-doped fibre amplifier (EDFA) before launch into the link.

Each channel was modulated using 64-QAM at a symbol rate of 47.5~GBd. The channels were spaced 50~GHz apart. The symbols were root-raised-cosine (RRC) shaped with a roll-off factor of 0.01. Transmitter compensation was applied, and the resulting waveforms were upsampled and loaded into a digital-to-analogue converter (DAC) with a sampling rate of 260~GSa/s.

The transmission link consisted of 30 spans of 100~km of pure-silica-core G.654 fibre, with a gain-flattening EDFA following each span to compensate the span loss and equalise the spectral power profile.

At the receiver, the signal was boosted by an EDFA to 27~dBm output power. The amplified signal was then divided into 20 branches using a power splitter; each branch passed through a narrowband bandpass filter, was amplified, and was fed into one of 20 dual-polarisation coherent receivers with 70-GHz photodiodes. The 80-channel, 80-GSa/s digital storage oscilloscope (DSO) with 36-GHz electrical bandwidth digitised the signal for offline processing. For the local oscillator (LO), a 25-GHz-spaced optical frequency comb with an interleaver (INT) and a 50-GHz-spaced AWG routed a single comb line to each of the 20 coherent receivers.

\section{Frequency Stitched Receiver}

\begin{figure}[ht!]
    \centering
    \begin{tikzpicture}
    \begin{axis}[
        width=\linewidth,
        height=0.8\linewidth,
        grid=both,
        label style={font=\footnotesize},
        xlabel={Total Launch Power (dBm)},
        ylabel={Total GMI Throughput (Gbps)},
        ymin=5700,
        legend cell align=left,
        legend style={font=\footnotesize, at={(0.03,0.03)}, anchor=south west}
    ]
        \addplot[uclblue, thick, mark=*, smooth] table[x=launch_dBm,y=total_GMI_edc] {data/64QAM_total.txt};
        \addlegendentry{EDC}
        \addplot[uclgreen, thick, mark=triangle, smooth] table[x=launch_dBm,y=total_GMI_perchan] {data/64QAM_total.txt};
        \addlegendentry{Per-Channel DBP}
        \addplot[uclred, thick, mark=diamond, smooth] table[x=launch_dBm,y=total_GMI_stitched] {data/64QAM_total.txt};
        \addlegendentry{Stitched DBP}

        \draw[uclblue, dashed] (axis cs:14,7254.4) -- (axis cs:20,7254.4);
        \draw[uclgreen, dashed] (axis cs:16,7414.3) -- (axis cs:20,7414.3);
        \draw[uclred, dashed] (axis cs:16,7647.6) -- (axis cs:20,7647.6);

        \draw[uclgreen, very thick, |->] (axis cs:18.5,7254.4) -- (axis cs:18.5,7414.3)
        node[pos=0,anchor=south east,align=right,inner ysep=0pt,style={font=\footnotesize\bfseries}]
        {160\\[-1pt]Gbps\\[-1pt](2.2\%)};

        \draw[uclred, very thick, |->] (axis cs:19,7254.4) -- (axis cs:19,7647.6)
        node[pos=0,anchor=south west,align=left,inner ysep=0pt,style={font=\footnotesize\bfseries}]
        {393\\[-1pt]Gbps\\[-1pt](5.4\%)};

\end{axis}
\end{tikzpicture}

\begin{tikzpicture}
    \begin{axis}[
        width=\linewidth,
        height=0.8\linewidth,
        grid=both,
        label style={font=\footnotesize},
        xlabel={Total Launch Power (dBm)},
        ylabel={Average SNR (dB)},
        ymin=9.5,
        legend cell align=left,
        legend style={font=\footnotesize, at={(0.03,0.03)}, anchor=south west, cells={align=left}}
    ]
        \addplot[uclblue, thick, mark=*, smooth] table[x=launch_dBm,y=avg_SNR_edc] {data/64QAM_total.txt};
        \addlegendentry{EDC}
        \addplot[uclgreen, thick, mark=triangle, smooth] table[x=launch_dBm,y=avg_SNR_perchan] {data/64QAM_total.txt};
        \addlegendentry{Per-Channel DBP}
        \addplot[uclred, thick, mark=diamond, smooth] table[x=launch_dBm,y=avg_SNR_stitched] {data/64QAM_total.txt};
        \addlegendentry{Stitched DBP}

        \draw[uclblue, dashed] (axis cs:14,12.3594) -- (axis cs:20,12.3594);
        \draw[uclgreen, dashed] (axis cs:16,12.6222) -- (axis cs:20,12.6222);
        \draw[uclred, dashed] (axis cs:16,13.0112) -- (axis cs:20,13.0112);

        \draw[uclgreen, very thick, |->] (axis cs:18.5,12.3594) -- (axis cs:18.5,12.6222)
        node[pos=1,anchor=south east,align=right,inner ysep=0pt,style={font=\footnotesize\bfseries}]
        {0.26 dB};

        \draw[uclred, very thick, |->] (axis cs:19,12.3594) -- (axis cs:19,13.0112)
        node[pos=0.5,anchor=south west,align=left,inner ysep=0pt,style={font=\footnotesize\bfseries}]
        {0.65 dB};

    \end{axis}
\end{tikzpicture}     \caption{Total GMI throughput (top) and average SNR (bottom) vs.\ total launch power for EDC, per-channel DBP, and full-field stitched DBP after 3000~km transmission.}
    \label{fig:throughput}
\end{figure}
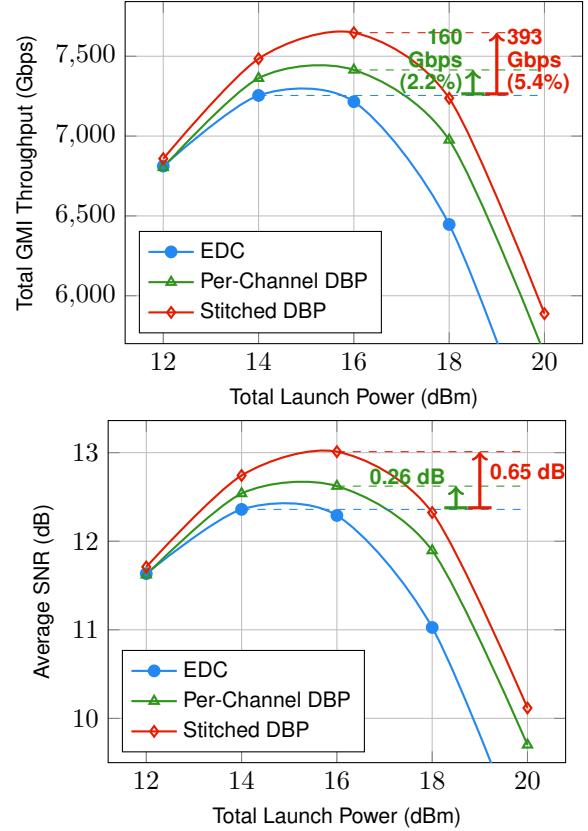

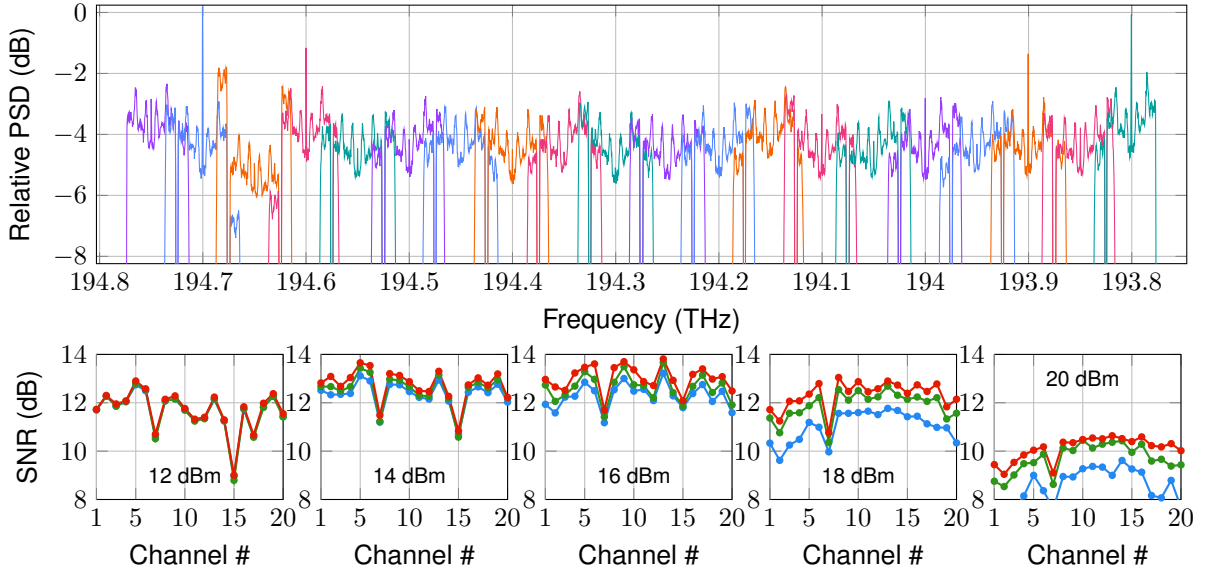
\begin{figure*}[ht!]
    \centering
    \begin{tikzpicture}
    \begin{axis}[
        name=topplot,
        width=\linewidth,
        height=5cm,
        ymin=-8,
        ymax=0,
        grid=both,
        xlabel={Frequency (THz)},
        ylabel={Relative PSD (dB)},
        cycle list name=uclcycle_bright,
        enlargelimits=0.03,
        x dir=reverse,
    ]
        \foreach \i in {1,...,20} {
            \addplot table[x=f\i,y=p\i] {data/spectrum_16dBm_iter2.txt};
        }
    \end{axis}

    \begin{groupplot}[
        group style={
            group size=5 by 1,
            horizontal sep=0.5cm,
            ylabels at=edge left,
        },
        width=4.05cm,
        height=3.5cm,
        every axis plot/.append style={mark size=1pt},
        grid=both,
        xlabel={Channel \#},
        ylabel={SNR (dB)},
ymin=8,ymax=14,
        xmin=1,xmax=20,
        extra x ticks={1},
    ]
        \nextgroupplot[
            at={($(topplot.south west)+(0,-1.2cm)$)},
            anchor=north west,
        ]
        \addplot[uclblue, thick, mark=*] table[x=channel,y=SNR_edc] {data/64QAM_12dBm.txt};
        \addplot[uclgreen, thick, mark=*] table[x=channel,y=SNR_perchan] {data/64QAM_12dBm.txt};
        \addplot[uclred, thick, mark=*] table[x=channel,y=SNR_stitched] {data/64QAM_12dBm.txt};
        \node at (axis cs:10,9) {\footnotesize 12~dBm};

        \nextgroupplot[
        ]
        \addplot[uclblue, thick, mark=*] table[x=channel,y=SNR_edc] {data/64QAM_14dBm.txt};
        \addplot[uclgreen, thick, mark=*] table[x=channel,y=SNR_perchan] {data/64QAM_14dBm.txt};
        \addplot[uclred, thick, mark=*] table[x=channel,y=SNR_stitched] {data/64QAM_14dBm.txt};
        \node at (axis cs:10,9) {\footnotesize 14~dBm};
        
        \nextgroupplot[
        ]
        \addplot[uclblue, thick, mark=*] table[x=channel,y=SNR_edc] {data/64QAM_16dBm.txt};
        \addplot[uclgreen, thick, mark=*] table[x=channel,y=SNR_perchan] {data/64QAM_16dBm.txt};
        \addplot[uclred, thick, mark=*] table[x=channel,y=SNR_stitched] {data/64QAM_16dBm.txt};
        \node at (axis cs:10,9) {\footnotesize 16~dBm};

        \nextgroupplot[
        ]
        \addplot[uclblue, thick, mark=*] table[x=channel,y=SNR_edc] {data/64QAM_18dBm.txt};
        \addplot[uclgreen, thick, mark=*] table[x=channel,y=SNR_perchan] {data/64QAM_18dBm.txt};
        \addplot[uclred, thick, mark=*] table[x=channel,y=SNR_stitched] {data/64QAM_18dBm.txt};
        \node at (axis cs:10,9) {\footnotesize 18~dBm};

        \nextgroupplot[
        ]
        \addplot[uclblue, thick, mark=*] table[x=channel,y=SNR_edc] {data/64QAM_20dBm.txt};
        \addplot[uclgreen, thick, mark=*] table[x=channel,y=SNR_perchan] {data/64QAM_20dBm.txt};
        \addplot[uclred, thick, mark=*] table[x=channel,y=SNR_stitched] {data/64QAM_20dBm.txt};
        \node at (axis cs:10,13) {\footnotesize 20~dBm};
    \end{groupplot}
\end{tikzpicture}
     \caption{Top: received optical spectrum at 16~dBm launch power. Bottom: per-channel SNR for each of the five launch powers tested (12--20~dBm), comparing EDC (blue), per-channel DBP (green), and full-field stitched DBP (red).}
    \label{fig:spectrum}
\end{figure*}

The 20 received sub-bands were first time-aligned using the overlap between adjacent sub-bands. The relative delay was estimated from the total received power, $|E_x|^2 + |E_y|^2$, which is insensitive to the inter-sub-band frequency offset. A coarse estimate was used to determine the integer-sample delay, followed by a fine estimate of the fractional delay from a linear phase ramp in the frequency domain. The adjacent sub-bands were then frequency-aligned by estimating the inter-sub-band spacing from the overlap region. This sub-picosecond timing alignment and sub-kHz frequency alignment were necessary to achieve a measurable full-field DBP gain. After time and frequency alignment, the sub-bands were stitched together to reconstruct the full 1-THz field, using raised-cosine weighting at the sub-band edges to smooth the transitions.

The DBP algorithm\footnote{\url{https://github.com/zceemja/labathon_ecoc2025/blob/12008e7e0aaca1b58ddc83505e010f9b2a82568e/functions.py\#L56}} was implemented using the split-step Fourier method (SSFM) to numerically solve the inverse Manakov equation~\cite{menyuk1987Nonlinear,ip2010Nonlinear}. The step size was chosen using adaptive local-error control following Sinkin et al.~\cite{sinkin2003Optimization}, with a target local error of $5\times10^{-9}$. Full-field DBP was performed at a sampling rate of 4~THz, while per-channel DBP used a 200-GHz sampling rate. The propagation model included both $\beta_2$ and $\beta_3$, estimated from the EDC stage: $\beta_2$ was obtained on a per-channel basis using the CMA-based dispersion estimate \cite[Eq.~(3)]{kuschnerov2009DSP}, and $\beta_3$ was obtained from a linear regression across the 20 channels. The resulting fibre parameters were $D = 20.1955$~ps/(nm$\cdot$km) and $S = 0.0628$~ps/(nm$^2\cdot$km) at the reference wavelength $\lambda_\text{ref} = 1543.13$~nm, taken as the centre of the 10 channel pairs. A parameter sweep for the 16-dBm result was used to determine the nonlinear coefficient $\gamma_\text{DBP}=0.45$~/W/km and a fibre loss of $\alpha_\text{DBP}=0.154$~dB/km.

For data recovery, the back-propagated signal was retimed to 2 samples/symbol, followed by LO carrier-frequency-offset removal. A data-aided LMS equaliser with carrier recovery in the loop was then applied, followed by a decision-directed LMS equaliser, again with carrier recovery in the loop. Performance was evaluated using the generalised mutual information (GMI).

\section{Results}

Figure~\ref{fig:throughput} shows the total GMI throughput and average SNR as a function of total launch power for EDC, per-channel DBP, and full-field stitched DBP. The peak throughput for EDC is 7254.4~Gbps at 14~dBm, for per-channel DBP 7414.3~Gbps at 16~dBm, and for full-field stitched DBP 7647.6~Gbps at 16~dBm. Full-field stitched DBP thus achieves a gain of 393~Gbps (5.4\%) over EDC, more than twice the 160~Gbps (2.2\%) gain achieved by per-channel DBP. In terms of average SNR, the peak values are 12.36~dB, 12.62~dB, and 13.01~dB for EDC, per-channel DBP, and stitched DBP, respectively, corresponding to gains of 0.65~dB and 0.26~dB for stitched DBP over EDC and per-channel DBP.

Figure~\ref{fig:spectrum} shows the reconstructed spectrum at the optimal launch power of 16~dBm at the top, and the per-channel SNR at the bottom for the measured launch powers. The stitched DBP consistently improves the SNR across all 20 channels, with the largest gains observed in channels 1, 4, 6, and 8--10. The average SNR gain over EDC per channel ranges from 0.3 to 1.3~dB at this launch power. As is visible in the received spectrum, the optical power is not uniform across sub-bands, meaning that the effective $\gamma P$ product used in the DBP model is inaccurate on a per-slice basis. A per-slice optimisation of the effective launch power used in the DBP model could reduce this mismatch and further improve the full-field compensation gain. This could be achieved by sweeping the $\gamma P$ product for each sub-band to maximise the recovered SNR. Furthermore, polarisation-mode dispersion (PMD) causes the field to decorrelate over long distances and limits the achievable gain unless it is explicitly accounted for in the back-propagation model~\cite{goroshko2015Fundamental,czegledi2017PMD}.

\begin{figure}[ht!]
    \centering
    \begin{tikzpicture}
    \begin{axis}[
        width=0.9\linewidth,
        height=4cm,
        grid=both,
        label style={font=\footnotesize, align=center},
        xlabel={Launch Power (dBm)},
        ylabel={Total number\\of DBP steps},
        legend cell align=left,
        legend style={font=\footnotesize, at={(1-0.03,0.5)}, anchor=east},
        ymode=log,
    ]

\addplot[uclgreen, thick, mark=triangle, smooth] table[x=launch_dBm,y=steps_perchan] {data/64QAM_total.txt};
        \addlegendentry{Per-Channel DBP}
        \addplot[uclred, thick, mark=diamond, smooth] table[x=launch_dBm,y=steps_stitched] {data/64QAM_total.txt};
        \addlegendentry{Stitched DBP}
    \end{axis}
\end{tikzpicture}     \caption{Number of SSFM steps vs.\ total launch power for per-channel and full-field stitched DBP.}
    \label{fig:steps}
\end{figure}
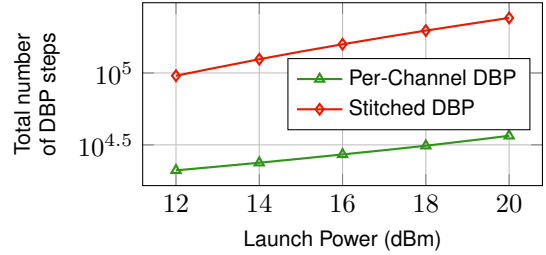

Figure~\ref{fig:steps} shows the number of SSFM steps required for convergence as a function of launch power. As expected, the number of steps increases with launch power due to the stronger nonlinearity. Full-field DBP requires approximately $5.8\times$ more steps than all per-channel DBP instances combined. This is on top of the 20$\times$ increase in bandwidth.

\section{Conclusion}

We have demonstrated full-field digital back-propagation over 3000~km with 1-THz receiver bandwidth, enabled by 20 synchronous coherent receivers, a frequency-comb local oscillator, and a propagation model including dispersion slope. Sub-picosecond timing alignment and sub-kHz frequency alignment between channels were necessary to observe full-field DBP gain. For a 20$\times$47.5~GBd, 50-GHz-spaced, PM-64QAM WDM signal, full-field DBP yielded a 393~Gbps (5.4\%) throughput gain over EDC, more than double the 160~Gbps (2.2\%) achieved by per-channel DBP.

\clearpage
\section{Acknowledgements}
The authors gratefully acknowledge all individuals and organisations that contributed to this research and supported the collaboration. This work was supported in part by the Engineering and Physical Sciences Research Council (EPSRC) under the programme grant “Transforming networks - building an intelligent optical infrastructure (TRANSNET)' (EP/R035342/1), and “Extremely Wideband Optical Fibre Communication Systems (EWOC)' (EP/W015714/1), and by the German Federal Ministry of Research, Technology and Space under the CELTIC-NEXT project SUSTAINET-Advance (16KIS2272). Robson A. Colares and Darli A. A. Mello are partially sponsored by CNPq grant \#405940/2022-0 and CAPES grant \#88887.954253/2024-00. Eric Sillekens' fellowship titled “Enabling Power Efficient Optical Communication through Novel Digital 
Signal Processing (EPIC DSP)” was supported by Department for Science, 
Innovation and Technology and the Royal Academy of Engineering under 
the Research Fellowships scheme, and Polina Bayvel by a Royal Society Research Professorship.
\printbibliography[]

@INPROCEEDINGS{deakin2024bandwidth,
  author={Deakin, Callum and Zang, Jizhao and Chen, Xi and Che, Di and Dallachiesa, Lauren and Stern, Brian and Fontaine, Nicolas K. and Papp, Scott},
  booktitle={{ECOC}}, 
  title={2.4-THz Bandwidth Optical Coherent Receiver Based on a Photonic Crystal Microcomb}, 
  year={2024},
  pages={766-769},
  address = {Frankfurt, Germany},
  }

@article{drayss2023Nonsliced,
  title = {Non-Sliced Optical Arbitrary Waveform Measurement ({{OAWM}}) Using Soliton Microcombs},
  author = {Drayss, Daniel and Fang, Dengyang and F{\"u}llner, Christoph and Lihachev, Grigory and Henauer, Thomas and Chen, Yung and Peng, Huanfa and {Marin-Palomo}, Pablo and Zwick, Thomas and Freude, Wolfgang and Kippenberg, Tobias J. and Randel, Sebastian and Koos, Christian},
  year = 2023,
  month = jul,
  journal = {Optica},
  volume = {10},
  number = {7},
  pages = {888--896},
  doi = {10.1364/OPTICA.484200}
}

@article{drayss2025Optical,
  title = {Optical Arbitrary Waveform Generation ({{OAWG}}) Using Actively Phase-Stabilized Spectral Stitching},
  author = {Drayss, Daniel and Fang, Dengyang and Sherifaj, Alban and Peng, Huanfa and F{\"u}llner, Christoph and Henauer, Thomas and Lihachev, Grigory and Schmitz, Lennart and Harter, Tobias and Freude, Wolfgang and Randel, Sebastian and Kippenberg, Tobias J. and Zwick, Thomas and Koos, Christian},
  year = 2025,
  month = sep,
  journal = {Light: Science \& Applications},
  volume = {14},
  number = {1},
  pages = {353},
  doi = {10.1038/s41377-025-01937-4}
}

@inproceedings{ali2018Impact,
  title = {The {{Impact}} of {{Dispersion Slope}} on {{Fiber Nonlinearity}} in {{Ultra-Wideband Optical Communication System}}},
  booktitle = {2018 {{European Conference}} on {{Optical Communication}} ({{ECOC}})},
  author = {Ali, Abdallah A. I. and Ferreira, F. and Al-Khateeb, M. and Charlton, D. and Laperle, C. and Ellis, A. D.},
  year = 2018,
  address = {Rome, Italy},
  publisher = {IEEE},
  doi = {10.1109/ECOC.2018.8535564}
}

@inproceedings{sohanpal2023Impact,
  title = {On the {{Impact}} of {{Frequency Variation}} on {{Nonlinearity Mitigation}} Using {{Frequency Combs}}},
  booktitle = {Optical {{Fiber Communication Conference}} ({{OFC}}) 2023},
  author = {Sohanpal, Ronit and Sillekens, Eric and Ferreira, Filipe M. and Killey, Robert I. and Bayvel, Polina and Liu, Zhixin},
  year = 2023,
  pages = {Th1F.3},
  publisher = {Optica Publishing Group},
  doi = {10.1364/OFC.2023.Th1F.3}
}

@article{kuschnerov2009DSP,
  title = {{{DSP}} for {{Coherent Single-Carrier Receivers}}},
  author = {Kuschnerov, M. and Hauske, F.N. and Piyawanno, K. and Spinnler, B. and Alfiad, M.S. and Napoli, A. and Lankl, B.},
  year = 2009,
  month = aug,
  journal = {Journal of Lightwave Technology},
  volume = {27},
  number = {16},
  pages = {3614--3622},
  doi = {10.1109/JLT.2009.2024963}
}

@article{dar2016Limits,
  title = {On the {{Limits}} of {{Digital Back-Propagation}} in {{Fully Loaded WDM Systems}}},
  author = {Dar, Ronen and Winzer, Peter J.},
  year = 2016,
  month = jun,
  journal = {IEEE Photonics Technology Letters},
  volume = {28},
  number = {11},
  pages = {1253--1256},
  doi = {10.1109/LPT.2016.2522969}
}

@article{menyuk1987Nonlinear,
  title = {Nonlinear Pulse Propagation in Birefringent Optical Fibers},
  author = {Menyuk, Curtis R.},
  year = 1987,
  month = feb,
  journal = {IEEE Journal of Quantum Electronics},
  volume = {23},
  number = {2},
  pages = {174--176},
  doi = {10.1109/JQE.1987.1073206}
}

@article{sinkin2003Optimization,
  title = {Optimization of the Split-Step Fourier Method in Modeling Optical-Fiber Communications Systems},
  author = {Sinkin, Oleg V. and Holzlohner, Ronald and Zweck, John and Menyuk, Curtis R.},
  year = 2003,
  month = jan,
  journal = {Journal of Lightwave Technology},
  volume = {21},
  number = {1},
  pages = {61--68},
  doi = {10.1109/JLT.2003.808628}
}

@article{czegledi2017PMD,
  title = {Digital Backpropagation Accounting for Polarization-Mode Dispersion},
  author = {Czegledi, Cristian B. and Liga, Gabriele and Lavery, Domanic and Karlsson, Magnus and Agrell, Erik and Savory, Seb J. and Bayvel, Polina},
  year = 2017,
  month = feb,
  journal = {Optics Express},
  volume = {25},
  number = {3},
  pages = {1903--1915},
  doi = {10.1364/OE.25.001903}
}

@article{civelli2025Twist,
  title = {A New Twist on Low-Complexity Digital Backpropagation},
  author = {Civelli, Stella and Jana, Debi Pada and Forestieri, Enrico and Secondini, Marco},
  year = 2025,
  journal = {Journal of Lightwave Technology},
  volume = {43},
  number = {10},
  pages = {4679--4692},
  doi = {10.1364/JLT.43.004679}
}

@inproceedings{fontaine2012228GHz,
  title = {228-{{GHz Coherent Receiver}} Using {{Digital Optical Bandwidth Interleaving}} and {{Reception}} of 214-{{GBd}} (856-{{Gb}}/s) {{PDM-QPSK}}},
  booktitle = {European {{Conference}} and {{Exhibition}} on {{Optical Communication}}},
  author = {Fontaine, Nicolas K. and Raybon, Greg and Guan, Binbin and Adamiecki, Andrew and Winzer, Peter J. and Ryf, Roland and Konczykowska, A. and Jorge, F. and Dupuy, J.-Y. and Buhl, L. L. and Chandrashekhar, S. and Delbue, R. and Pupalaikis, P. and Sureka, A.},
  year = 2012,
  pages = {Th.3.A.1},
  address = {Amsterdam},
  doi = {10.1364/ECEOC.2012.Th.3.A.1},
}

@article{galdino2017limits,
  title = {On the Limits of Digital Back-Propagation in the Presence of Transceiver Noise},
  author = {Galdino, Lidia and Semrau, Daniel and Lavery, Domanic and Saavedra, Gabriel and Czegledi, Cristian B. and Agrell, Erik and Killey, Robert I. and Bayvel, Polina},
  year = 2017,
  month = feb,
  journal = {Optics Express},
  volume = {25},
  number = {4},
  pages = {4564},
  doi = {10.1364/OE.25.004564}
}

@inproceedings{goroshko2015Fundamental,
  title = {Fundamental {{Limitations}} of {{Digital Back Propagation}} Due to {{Polarization Mode Dispersion}}},
  booktitle = {Asia {{Communications}} and {{Photonics Conference}} 2015},
  author = {Goroshko, Kseniia and Louchet, Hadrien and Richter, Andre},
  year = 2015,
  pages = {ASu3F.5},
  address = {Hong Kong},
  doi = {10.1364/ACPC.2015.ASu3F.5},
}

@article{ip2010Nonlinear,
  title = {Nonlinear {{Compensation Using Backpropagation}} for {{Polarization-Multiplexed Transmission}}},
  author = {Ip, Ezra},
  year = 2010,
  month = mar,
  journal = {Journal of Lightwave Technology},
  volume = {28},
  number = {6},
  pages = {939--951},
  doi = {10.1109/JLT.2010.2040135}
}

@inproceedings{shi2017246a,
  title = {246 {{GHz Digitally Stitched Coherent Receiver}}},
  booktitle = {Optical {{Fiber Communication Conference}}},
  author = {Shi, Kai and Sillekens, Eric and Thomsen, Benn C.},
  year = 2017,
  pages = {M3D.3},
  address = {Los Angeles, California},
  doi = {10.1364/OFC.2017.M3D.3},
}

\vspace{-4mm}

\end{document}